\documentstyle[twocolumn,aps]{revtex}
\begin{document}
\draft
\title{Observation of anomalous single-magnon scattering\\
in half-metallic ferromagnets by chemical pressure control}
\author{T. Akimoto} 
\address{Department of Crystalline Materials Science, Nagoya University
, Nagoya 464-8603, Japan}
\author{Y. Moritomo\footnote{to whom correspondence should be addressed} and A. Nakamura}
\address{Center for Integrated Research in Science and Engineering,\\ Nagoya University, Nagoya 464-8601}
\author{N. Furukawa}
\address{Department of Physics, Aoyama Gakuin University, Setagaya, Tokyo 157-8572, Japan}
\date{Received 17 April 2000}
\maketitle

\begin{abstract}
Temperature variation of resistivity $\rho$ and specific heat $C$ have been measured for prototypical half-metallic ferromagnets, $R_{0.6}$Sr$_{0.4}$MnO$_{3}$, with controlling the one-electron bandwidth $W$.
We have found variation of the temperature scalings in $\rho$ from $\sim T^2$ ($R$ = La, and Nd) to $\sim T^3$ ($R$ = Sm), and have interpreted the $T^3$-law in terms of the {\it anomalous} single-magnon scattering (AMS) process in the half-metallic system.

\end{abstract}
\pacs{72.90.+y,72.60.+g}

\narrowtext
So far, several ferromegnetic metallic systems, such as CrO$_2$\cite{Soulen} and doped manganites,\cite{Soulen,Park} are found to be half-metallic.
Especially, Park {\it et al.}\cite{Park} have directly confirmed the half-metallic electronic structure of a doped manganite by means of the spin-resolved photo-emission spectroscopy.
The doped manganites, $R_{1-x}A_{x}$MnO$_{3}$ ($R$ and $A$ being the trivalent rare-earth and divalent alkaline-earth ions, respectively), has a distorted perovskite structure, and hence we can control the one-electron bandwidth $W$ of the $e_g$-carriers by changing the averaged ionic radius $r_{\rm A}$ of the perovskite $A$-site ({\it chemical pressure}).\cite{Hwang,Moritomo}
Thus, the doped manganite is the unique half-metallic system whose material parameters, e.g., $W$, can be tuned by the chemical pressure.

In the half-metallic system, such as doped manganites, single-magnon scattering process ($\sim T^2$\cite{Mannari}) is prohibited, since only one spin channel is metallic.
So, Kubo and Ohata\cite{Kubo} have calculated effect of the two-magnon scattering process on resistivity $\rho$, and have derived $T^{9/2}$ dependence.
The $\rho-T$ curve of most of the doped manganites, however, shows $T^2$ behavior in a wide temperature range in the ferromagnetic phase.\cite{Urushibara,Snyder}
This unexpected temperature dependence has been ascribed to the electron-electron scattering.\
Jaime {\it et al.}\cite{Jaime} have carefully measured resistivity of a single crystalline La$_{0.67}$(Pb,Ca)$_{0.33}$MnO$_{3}$, and have observed $T^{1.5}$ dependence below $\sim$ 50 K.
A similar deviation from the $T^2$-law has been reported by Furukawa.\cite{Furukawa}
Recently, Furukawa\cite{Furukawa} has predicted that an {\it anomalous} single-magnon scattering (AMS) process gives $\sim (T/D_{\rm s})^{3}$ dependence, where $D_{\rm s}$ is the spin-wave stiffness coefficient, in a half-metallic system
This is because the inverse lifetime of the majority spin carriers is proportional to the density of state (DOS) of the minority carrier band as well as the magnon density. 
If this model is correct, we expect that the $T^3$-component becomes dominant with the decrease of $D_{\rm s}$ ($\propto W$).

In this Letter, we have systematically investigated resistivity and specific heat for the half-metallic manganites, $R_{0.6}$Sr$_{0.4}$MnO$_{3}$, with changing rare earth element $R$.
We have found that the scattering process changes from at the electron-electron-type ($\rho \sim T^2$; $R$ =La and Nd), the single-magnon-type ($\rho \sim T^3$; $R$ = Sm and Nd$_{0.8}$Tb$_{0.2}$) to the two-magnon-type ($\rho \sim T^{9/2}$; $R$ = Nd$_{0.6}$Tb$_{0.4}$) with the decrease of $D_{\rm s}$. 

Crystals of $R_{0.6}$Sr$_{0.4}$MnO$_{3}$ ($R$ = La, Nd, Nd$_{0.8}$Tb$_{0.2}$, Sm and Nd$_{0.6}$Tb$_{0.4}$) were grown by the floating-zone method at a feeding speed of 10 - 15 mm/h.
A stoichiometric mixture of commercial La$_2$O$_3$, Nd$_2$O$_3$, Sm$_2$O$_3$, Tb$_4$O$_7$, SrCO$_3$, Mn$_3$O$_4$ powder was ground and calcined two times at 1350 $^\circ$C for 24 h.
Then, the resulting powder was pressed into a rod with a size of $\phi$5 $\times$ 60 mm and sintered at 1350 $^\circ$C  for 24 h.
The ingredient could be melted congruently in a flow of air.
Large crystals, typically 4 mm in diameter and 20 mm in length, were obtained.
Powder x-ray diffraction measurements at room temperature and Reitveld analysis indicate that the crystals were single phase without detectable impurity.
Except for La$_{0.6}$Sr$_{0.4}$MnO$_{3}$ (Rhombohedral: $R\bar{3}c$, $Z$ = 2), the crystal symmetry is orthorhombic ($Pbnm$; Z = 4).
We further carefully investigated the low temperature powder patterns, and observed no trace of the phase mixing.
To determine the Curie temperature $T_{\rm C}$, temperature dependence of the magnetization $M$ was measured under a field of $\mu_{0}H$ = 0.5 T after cooling down to 5 K in zero field (ZFC), and is shown in the inset of Fig.\ref{fig1}.
All the samples investigated become ferromagnetic at low temperature (downward arrow in the inset) with saturation magnetization of $\sim$ 3.5 $\mu_{\rm B}$.
$T_{\rm C}$ was determined from the inflection point of the $M-T$ curve.
We plotted thus obtained $T_{\rm C}$ in Fig.\ref{fig1} against $r_{\rm A}$.
With the decrease of $r_{\rm A}$, $T_{\rm C}$ decreases from $\approx$ 360 K for $R$ = La to $\approx$ 85 K for $R$ = Nd$_{0.6}$Tb$_{0.4}$.

First, we have carefully investigated temperature variation of $\rho$ in $R_{0.6}$Sr$_{0.4}$MnO$_{3}$, and have found a systematic variation of the scaling relation from $\sim T^2$, $\sim T^3$ to $\sim T^{9/2}$ with the decrease of $r_{\rm A}$.
For four-probe resistivity measurements, the crystal was cut into a rectangular shape, typically of 3$\times$1$\times$0.5 mm$^{3}$, and electrical contacts were made with a heat-treatment-type silver paint.
In Fig.~\ref{fig2} are shown temperature variation of resistivity of (a) $R$ = La and Nd, (b) $R$ = Sm and (c) $R$ = Nd$_{0.6}$Tb$_{0.4}$.
The $T^2$ and $T^{9/2}$ dependence can be ascribed to the electron-electron scattering\cite{t2} and to the two-magnon scattering\cite{Kubo}, respectively.
We tentatively ascribed the $T^3$-law to the AMS proposed by Furukawa.\cite{Furukawa}

To qualitatively estimate the contribution of the respective scattering processes, we have fitted the $\rho-T$ curve in the low temperature region with the following equation:
\begin{equation}
\rho(T) - \rho_0 = A_2T^{2} + A_3T^{3} + A_{9/2}T^{9/2},
\label{eq1}
\end{equation}
where $\rho_0$ is residual resistivity estimated by extrapolation of the $\rho-T$ curve.
The temperature range used for the fitting is far below $T_{\rm C}$: temperature range is below $\sim$ 100 K for $R$ = La and Nd ($T_{\rm C} \geq$ 300 K) and below $\sim$ 40 K for $R$ = Nd$_{0.8}$Tb$_{0.2}$ and Sm ($T_{\rm C} \geq$ 120 K). We estimated the relative weight of the respective scattering processes with use of $A_{2}$, $A_{3}$ and $A_{9/2}$:
\begin{equation}
a_2 = \int_0^{T_{\rm max}}A_2T^{2}dt/\int_0^{T_{\rm max}}[\rho(T) - \rho_0]dt,
\label{eq2}
\end{equation}
\begin{equation}
a_3 = \int_0^{T_{\rm max}}A_3T^{3}dt/\int_0^{T_{\rm max}}[\rho(T) - \rho_0]dt 
\label{eq3}
\end{equation}
and 
\begin{equation}
a_{9/2} = \int_0^{T_{\rm max}}A_{9/2}T^{9/2}dt/\int_0^{T_{\rm max}}[\rho(T) - \rho_0]dt, 
\label{eq4}
\end{equation}
where $T_{\rm max}$ ( = 30 K) is the cut-off temperature.

As described in the introduction, the dominating scattering process of the half-metallic ferromagnet is expected to change with $D_{\rm s}$.
To evaluate the variation of the $D_{\rm s}$ with chemical pressure, we have measured low temperature specific heat $C$ by means of the relaxation method.
The total specific heat is composed of four parts, namely, contribution from conduction electrons, lattice, spin-waves and localized 4$f$-electrons:
\begin{equation}
C = C_{\rm ele} + C_{\rm spin} + C_{\rm lat} + C_{\rm Sch},
\end{equation}
where $C_{\rm ele}$, $C_{\rm spin}$, $C_{\rm lat}$ and $C_{\rm Sch}$ are expressed as $\gamma T$, $\alpha T^{3/2}$, $\beta T^{3}$+$\beta_{1} T^{5}$ and $\delta T^{-2} \cdot$e$^{\Delta E/k_{\rm B}T}$(1+e$^{\Delta E/k_{\rm B}T}$)$^{-2}$, respectively.
The additional second term ($\beta_{1} T^{5}$\cite{Woodfield}) in $C_{\rm lat}$ is needed to improve the fitting in the higher temperature region.
Note that contribution of the second term is small ($\leq$ 1 - 18 \% at 10 K).
We show the best-fitted results in Fig.\ref{fig3}, in which the Schottoky component $C_{\rm Sch}$ was subtracted for convenience of explanation.
Inset shows magnon component of the specific heat ($C$ -  $C_{\rm Sch}$ - $C_{\rm ele}$ - $C_{\rm lat}$) of $R_{0.6}$Sr$_{0.4}$MnO$_{3}$ against $T^{3/2}$.
As indicated by straight lines, the component ($\alpha T^{3/2}$) systematically increases with the decrease of $r_{\rm A}$ ($T_{\rm C}$).
Thus obtained parameters, {\it i.e.}, $\alpha$, $\beta$, $\beta_1$, $\gamma$, $\delta$ and $\Delta E$, are listed in table~\ref{table}.
The parameters of La$_{0.6}$Sr$_{0.4}$MnO$_3$ are nearly the same as those of La$_{0.7}$Sr$_{0.3}$MnO$_3$ ceramics.\cite{Woodfield}
We have calculated the $D_{\rm s}$-value from $\alpha$ ( $\equiv$ 0.113 $k_{\rm B}^{5/2}$/$D_{\rm s}^{3/2}$), and plotted them in Fig.\ref{fig4}(a) against $T_{\rm C}$.
In the same figure, we also plotted the $D_{s}$-values (open squares) of La$_{0.7}$Sr$_{0.3}$MnO$_{3}$ and La$_{0.8}$Sr$_{0.2}$MnO$_{3}$ determined by the inelastic neutron scattering measurements\cite{Endoh}.
Presently obtained $D_{\rm s}$-values are consistent with them, even though each value has a large error bar.
The reduced $D_{\rm s}$ can be ascribed to the reduced $W$-value of the low-$T_{\rm C}$ compound.\cite{Zener} 

To investigate the interrelation between the $D_{\rm s}$-value and the dominating scattering processes, we plotted the relative weight of the resistivity components, namely, $a_2$ (eq.\ref{eq2}), $a_3$ (eq.\ref{eq3}) and $a_{9/2}$ (eq.\ref{eq4}), in the lower panel of Fig.\ref{fig4}.
With the decrease of $D_{\rm s}$ ($T_{\rm C}$), the dominating term changes from $\sim T^2$ ($R$ = La and Nd), $\sim T^3$ ($R$ = Nd$_{0.8}$Tb$_{0.2}$ and Sm) to $\sim T^{9/2}$ ($R$ =Nd$_{0.6}$Tb$_{0.4}$).

Now, let us discuss the origin of the close correlation between $D_{\rm s}$ and the dominating resistivity term.
Even in the half-metallic ferromagnet, the spin fluctuation at a finite temperature induces the DOS in the minority spin band.
In this situation, the single-magnon scattering process would be allowed, and would give $\sim (T/D_{\rm s})^3$ dependence: the scattering intensity is proportional to the DOS of the minority carrier band as well as the magnon density.
Here, we should emphasize that this AMS process is enhanced with the decrease of $D_{\rm s}$.
This is consistent with the experimental results: the dominating resistivity component changes from $\sim T^2$ (electron-electron scattering\cite{t2}) to $\sim T^3$, when the $D_{\rm s}$-value decreases below $\sim$ 50 meV\AA$^2$ (see Fig.\ref{fig4}).
Therefore, the presently observed $\sim T^3$-law of Sm$_{0.6}$Sr$_{0.4}$MnO$_3$ and (Nd$_{0.8}$Tb$_{0.2}$)$_{0.6}$Sr$_{0.4}$MnO$_3$ can be ascribed to the AMS.
With a further decrease of $D_{\rm s}$, the two-magnon scattering process, which gives $\sim (T/D_{\rm s})^{9/2}$ dependence\cite{Kubo}, is expected to become dominant.
Actually, we observed $\sim T^{9/2}$ dependence in (Nd$_{0.6}$Tb$_{0.4}$)$_{0.6}$MnO$_3$ ($D_{\rm s} \sim$ 20 meV\AA$^2$).
The low temperature deviation from the $T^{9/2}$-law observed in Fig.\ref{fig2}(c) may be ascribed to the single-magnon scattering process.

In summary, we have systematically investigated the low temperature resistivity $\rho$ and specific heat $C$ for the prototypical half-metallic ferromagnet, $R_{0.6}$Sr$_{0.4}$MnO$_{3}$.
With the decrease of $D_{\rm s}$, the low temperature resistivity change from the $T^{2}$-type to $T^{3}$-type due to the AMS process.
Here, let us comment on the transport properties of the other half-metallic ferromagnets, {\it i.e.}, CrO$_2$\cite{Soulen,Karlheinz} and Sr$_2$FeMoO$_6$\cite{Kobayashi,FeMo}.
The $\rho-T$ curves of both the compounds seriously deviate from the $T^2$-law in the low temperature region below $\sim$ 100 K.\cite{Suzuki,MoritomoFeMo}
Such a deviation is usually ascribed to the electron-phonon scattering, which gives $\sim T^5$-depedence at low temperature.
However, the AMS is another candidate for the deviation in the half-metallic system.\\

T. A. acknowledges the financial support from JSPS Research Fellowship for Young Scientists.
This work was supported by a Grant-In-Aid for Scientific Research from the Ministry of Education, Science, Sports and Culture, and from Daiko foundation.

\begin{table}
\caption{Parameters obtained from the specific heat measurements of $R_{0.6}$Sr$_{0.4}$MnO$_{3}$. Total specific heat is composed of four parts, namely, contribution from conduction electrons ($C_{\rm ele}$), lattice ($C_{\rm lat}$), spin-waves ($C_{\rm spin}$) and localized 4$f$-electrons ($C_{\rm Sch}$). $C_{\rm ele}$, $C_{\rm spin}$, $C_{\rm lat}$ and $C_{\rm Sch}$ are expressed as $\gamma T$, $\alpha T^{3/2}$, $\beta T^{3}$+$\beta_{1} T^{5}$ and $\delta T^{-2} \cdot$e$^{\Delta E/k_{\rm B}T}$(1+e$^{\Delta E/k_{\rm B}T}$)$^{-2}$, respectively.}
\label{table}
\begin{tabular}{|c||cccccc|}
$R$&$\gamma$&$\alpha$&$\beta$&$\beta_{1}$&$\delta$&$\Delta E$\\
&mJ/K$^{2}$mol&mJ/K$^{5/2}$mol&10$^{-2}$mJ/K$^4$mol&10$^{-4}$mJ/K$^6$mol&10$^{4}$mJ/K$^{-1}$mol&K\\
\hline
La&3.83&1.6&9.14&1.96&---&---\\
Nd&3.77&2.9&14.3&1.96&2.79&7.7\\
Nd$_{0.8}$Tb$_{0.2}$&3.76&10.0&18.7&0.21&4.92&11.6\\
Sm&3.65&18.5&13.2&0.19&6.20&13.8\\
Nd$_{0.6}$Tb$_{0.4}$&3.06&14.4&17.7&0.20&6.40&13.6\\
\end{tabular}

\end{table}

\begin{figure}
\caption{Interrelation between Curie temperature $T_{\rm C}$ and  average ionic radius $r_{\rm A}$ of perovskite $A$-site for $R_{0.6}$Sr$_{0.4}$MnO$_{3}$, where $R$ is a trivalent rare-earth ion. Inset shows temperature dependence of the magnetization $M$ measured at 0.5 T after cooling down to 5 K in zero field (ZFC). $T_{\rm C}$ was determined from the inflection point of the $M-T$ curve.}
\label{fig1}
\end{figure}

\begin{figure}
\caption{Temperature dependence of resistivity: (a) La$_{0.6}$Sr$_{0.4}$MnO$_{3}$ and Nd$_{0.6}$Sr$_{0.4}$MnO$_{3}$, (b) Sm$_{0.6}$Sr$_{0.4}$MnO$_{3}$ and (c) (Nd$_{0.6}$Tb$_{0.4}$)$_{0.6}$Sr$_{0.4}$MnO$_{3}$. Residual resistivity $\rho_0$ was subtracted. Solid curves are the best fitted results with equation: $\rho$($T$) - $\rho_{\rm 0}$ = $A_{2}T^{2}$ + $A_{3}T^{3}$ + $A_{9/2}T^{9/2}$.}
\label{fig2}
\end{figure}

\begin{figure}
\caption{Temperature dependence of the subtracted specific heat $C$ - $C_{\rm Sch}$ of $R_{0.6}$Sr$_{0.4}$MnO$_{3}$ ($R$ = La and Nd). $C_{\rm Sch}$ is the Schottky component. Solid curves are the best fitted results taking into account of contributions from electrons, spin-waves and lattice: $C$ - $C_{\rm Sch}$ = $\gamma T$ + $\alpha T^{3/2}$ + $\beta T^{3}$ + $\beta_{1} T^{5}$. Inset shows magnon contribution to the specific heat ($C$ -  $C_{\rm Sch}$ - $C_{\rm ele}$ - $C_{\rm lat}$ = $\alpha T^{3/2}$) of $R_{0.6}$Sr$_{0.4}$MnO$_{3}$ against $T^{3/2}$. Straight lines are results of the least-square fittings.}
\label{fig3}
\end{figure}

\begin{figure}
\caption{(a) Spin-wave stiffness coefficient $D_{\rm s}$ (open circles) estimated from low temperature specific heat of $R_{0.6}$Sr$_{0.4}$MnO$_{3}$. $T_{\rm C}$ is the Curie temperature. Open squares are the data obtained by the inelastic neutron scattering (Cited from Ref.[13]). (b) Relative weight of the resistivity components, {\it i.e.}, $a_2$, $a_3$ and $a_{9/2}$., with $\sim T^2$, $\sim T^3$ and $\sim T^{9/2}$ dependence.
$a_2$, $a_3$ and $a_{9/2}$ are defined by $\int_0^{T_{\rm max}}A_2T^{2}dt/\int_0^{T_{\rm max}}[\rho(T) - \rho_0]dt$, $\int_0^{T_{\rm max}}A_3T^{3}dt/\int_0^{T_{\rm max}}[\rho(T) - \rho_0]dt$ and $\int_0^{T_{\rm max}}A_{9/2}T^{9/2}dt/\int_0^{T_{\rm max}}[\rho(T) - \rho_0]dt$, respectively. $T_{\rm max}$ (= 30 K) is the cut-off temperature. Each components, {\it i.e.}, $A_2T^2$, $A_3T^3$ and $A_{9/2}T^{9/2}$, are determined by fitting of the low temperature resistivity.}
\label{fig4}
\end{figure}


\begin{references}
\bibitem{t2} Another probable origin for the $T^2$-law is the conventional one-magnon scatteroing process due to the thermally-destoried half-metalic state. Actually, the low temperature deviation from the $T^2$-law has been ascribed to the AMS scattering process in the half-metallic state.\cite{Furukawa}
\bibitem{FeMo} Sr$_2$FeMoO$_6$ has a ferrimagnetic structure, and has the acoustic and optical branches of the spin-wave. Dispersion ($E$($q$) = $D_{\rm s}^- q^2$) of the former branch is equivalent to that of ferromagnet.
\end{references}
\end{document}